\documentclass{aastex631}

\usepackage{appendix}
\usepackage{graphicx}

\begin{document}

\title{Highly depleted alkali metals in Jupiter’s deep Atmosphere}

\author[0000-0003-1898-8080]{Ananyo Bhattacharya}
\correspondingauthor{Ananyo Bhattacharya}
\email{ananyo@umich.edu}
\affiliation{University of Michigan, Ann Arbor}

\author{Cheng Li}
\affiliation{University of Michigan, Ann Arbor}

\author{Sushil K. Atreya}
\affiliation{University of Michigan, Ann Arbor}

\author{Paul G. Steffes}
\affiliation{Georgia Institute of Technology}

\author{Steven M. Levin}
\affiliation{NASA Jet Propulsion Laboratory}

\author{Scott J. Bolton}
\affiliation{Southwest Research Institute}

\author{Tristan Guillot}
\affiliation{Universite C´ote d’Azur}

\author{Pranika Gupta}
\affiliation{University of Michigan}

\author{Andrew P. Ingersoll}
\affiliation{California Institute of Technology}

\author{Jonathan I. Lunine}
\affiliation{Cornell University}

\author{Glenn S. Orton}
\affiliation{NASA Jet Propulsion Laboratory}

\author{Fabiano A. Oyafuso}
\affiliation{NASA Jet Propulsion Laboratory}

\author{J. Hunter Waite}
\affiliation{Southwest Research Institute}
\affiliation{Waite Science LLC}

\author{Amadeo Belloti}
\noaffiliation

\author{Michael H. Wong}
\affiliation{University of California, Berkeley}



\begin{abstract}

Water and ammonia vapors are known to be the major sources of spectral absorption at pressure levels observed by the microwave 
radiometer (MWR) on Juno. However, the brightness temperatures and limb darkening observed by the MWR at its longest wavelength 
channel of 50 cm (600 MHz) in the first 9 perijove passes indicate the existence of an additional source of opacity in the deep 
atmosphere of Jupiter (pressures beyond 100 bar). The absorption properties of ammonia and water vapor, and their relative 
abundances in Jupiter’s atmosphere do not provide sufficient opacity in the deep atmosphere to explain the 600 MHz channel 
observation. Here we show that free electrons due to the ionization of alkali metals, i.e. sodium, and potassium, with sub-solar 
metallicity, [M/H] (log based 10 relative concentration to solar) in the range of [M/H] = -2 to [M/H] = -5 can provide the 
missing source of opacity in the deep atmosphere. If the alkali metals are not the source of additional opacity in the MWR data, 
then their metallicity at 1000 bars can only be even lower.  This upper bound of -2 on the metallicity of the alkali metals 
contrasts with the other heavy elements -- C, N, S, Ar, Kr, and Xe -- which are all enriched relative to their solar abundances 
having a metallicity of approximately +0.5.    

\end{abstract}

\keywords{Solar System (1528) -- Chemical abundances(224) -- Jupiter(873) -- Extrasolar gaseous giant planets(509)}


\section{Introduction} 

The alkali metals sodium and potassium have been previously detected in the atmospheres of hot Jupiters and a super-Neptune 
together with lithium [\cite{chen2018gtc}] in the latter. The detections show a large range of abundances from highly substellar 
to super-stellar values [\cite{welbanks2019mass}, \cite{demory2011high}]. Alkali metal abundances are important in understanding 
the formation of  hot Jupiters and represent a bridge between the refractory and volatile elements, which in molecular form seed 
the growth of planets. Obtaining the abundance of alkali metals in Jupiter can potentially serve as a first constraint on the 
ratio of rocky to icy material in the interior of the solar system’s largest planet when combined with the elemental and 
molecular abundances provided by the Galileo Probe Mass Spectrometer (GPMS) [\cite{atreya1999comparison}, \cite{wong2004updated}, 
\cite{atreya2019origin}] and Juno constraints on water [\cite{li2020water}]. Here we derive observationally based abundances of 
alkali metals in Jupiter’s atmosphere to  determine whether they are enriched relative to solar like the other heavy elements or 
depleted.\\

To obtain these abundances requires knowing the deep structure of Jupiter’s atmosphere. The shallower part of Jupiter’s 
atmosphere has been previously investigated at microwave frequencies by the Very Large Array (VLA) telescope [\cite{de2003vla}, 
\cite{de2019jupiter}]. VLA probes Jupiter at frequencies in the range of 74 MHz to 50 GHz [\cite{de2019jupiter}]. However, 
confusion from Jupiter’s powerful synchrotron radiation does not allow VLA to observe Jupiter’s atmosphere below 5 GHz 
[\cite{de2003vla}], limiting its reach to less than 5 bars, leaving the deep atmosphere of Jupiter inaccessible from microwave 
and radio frequency observatories from Earth. The orbit of Juno and the spin of the spacecraft allow the spacecraft to make 
observations at low frequencies, i.e. 0.6 GHz and 1.2 GHz, by avoiding the energetic electron belts around Jupiter from its field 
of view. Access to greater depths allows for the investigation of bulk elemental abundances of N and O in Jupiter 
[\cite{janssen2017mwr}, \cite{bolton2017jupiter}, \cite{steffes2017high}].\\

The Microwave Radiometer (MWR) instrument onboard the Juno orbiter is a passive radiometer that is designed to measure the 
internal heat emitted by Jupiter’s atmosphere at six different frequencies ranging from 0.6 GHz to 22 GHz 
[\cite{janssen2017mwr}]. The brightness temperature measured by MWR at these frequencies sounds different levels of Jupiter’s 
atmosphere corresponding to pressures from 0.3 bar to 250 bar [\cite{janssen2017mwr}]. In addition, the highly inclined polar 
orbit and rotation of the Juno spacecraft aided in the high spatial resolution necessary for probing Jupiter’s atmosphere at 
various latitudes [\cite{bolton2017jupiter}].\\

Previous analysis of the MWR data at the 0.6 GHz found an unanticipated limb-darkening signal, which cannot be explained by 
nominal absorbers such as ammonia and water [\cite{li2020water}]. Based on investigation of thermodynamic models of Jupiter’s 
deep atmosphere between 50 bar and 1 kbar 
 [\cite{fegley1994chemical}, \cite{weidenschilling1973atmospheric}] we conjecture that the free electrons from thermally ionzied 
 alkali metals may provide the missing opacity. Alkali metals are expected to undergo condensation to form clouds in the deep 
 atmosphere [\cite{visscher2006atmospheric}, \cite{morley2012neglected}]. Na$_{2}$S and KCl are the first chemical species to 
 condense in the above pressure range and thereby act as a sink for atomic sodium and potassium [\cite{fegley1994chemical}]. 
 Furthermore, high-temperature environments cause alkali metals to undergo ionization due to their low ionization energies 
 [\cite{bagenal2007jupiter}]. Density and temperature play a role in governing the electron densities according to the Saha 
 ionization equation (Eq. 2). Electrons generated from alkali metal ionization act as a source of absorption at microwave 
 frequencies that could affect the brightness temperatures at the 0.6 GHz frequency channel. Therefore, the objective of this 
 study is to determine the alkali metal abundance in the deep atmosphere of Jupiter.\\

To facilitate comparison of our results on alkali metals with those of the extrasolar planets we express the abundances of non-
hydrogen and helium elements using astronomical terminology, e.g., metallicity. The metallicity (\emph{[M/H]}) of an element is 
the logarithm of the ratio of elemental abundance in a system to the stellar (or solar, for the solar system) elemental 
abundance. Generally, the metallicity of a star is defined in terms of the ratio of the number of Fe atoms to the number of 
hydrogen atoms. Here we define the metallicity in terms of alkali metal abundance in Jupiter to that of Sun e.g. for potassium, 
\emph{[K/H]} = log$_{10}$(\emph{N$_{K}$/N$_{H}$})$_{Jupiter}$ - log10(\emph{N$_{K}$/N$_{H}$})$_{Sun}$. For the giant planets, 
iron and silicon is not measurable, emphasizing the importance of proxy indicators such as the alkali metals along with other 
elements measured by Galileo probe. \\

\section{Methods} \label{sec:methods}
Brightness temperatures from 9 perijoves i.e. PJ 1,3-9, 12 have been taken into consideration for this article. Variations in 
brightness temperatures have been observed across the planetocentric latitudes from pole-to-pole at 0.6 and 1.2 GHz channels. 
These variations can be attributed to various sources of origin from the atmosphere and space environment. The most important 
sources of the observed variability are (i) changes in atmospheric structure and composition, (ii) Jupiter’s synchrotron 
radiation in the microwave band, and (iii) variation in acceleration due to gravity due to the non-spherical shape of Jupiter. 
The latter sources, i.e. synchrotron and gravity need to be taken into account for proper interpretation of MWR observations. It 
will aid in investigating the true variability in Jupiter’s deep atmosphere.\\

The contribution of Jupiter’s gravity can be corrected by taking into account the non-spherical shape of Jupiter. Brightness 
temperatures are corrected using a gravity correction factor defined as the ratio of theoretical \emph{T$_{b}$} at a given 
latitude to that at the equator of Jupiter taking into consideration the acceleration due to gravity at the latitude. Thereby, it 
transforms the Juno observations at each latitude for equatorial gravity, which effectively removes variation in \emph{T$_{b}$} 
due to changes in Jupiter’s gravity from the equator to the poles.\\

Energetic electrons in Jupiter’s space environment contribute to the synchrotron radiation [\cite{de2003vla}, 
\cite{levin2001modeling}, \cite{santos2017first}]. The signature of the emission is observed in MWR data across all the perijoves 
which leads to anomalous changes in \emph{T$_{b}$}. Data at extremely high latitudes are polluted by synchrotron emission and 
thus, remain of no use for investigating Jupiter's deep atmosphere. Therefore, we only consider the MWR data between -60 to 60 
deg. latitude. The correction for synchrotron and other sources of anomalous \emph{T$_{b}$} is done by filtering the data at 0.6 
and 1.2 GHz for each perijove. The process is carried out by sorting the deviations of \emph{T$_{b}$} from the least value of 
T$_{b}$ in a group and removing the values greater than a filter cutoff temperature of the order of 2 K.\\
\section{Results} \label{sec:results}

\subsection{Sources of Microwave Opacity}

The weighting function of Jupiter’s atmospheric absorption and emission at a given microwave frequency determines the 
contribution of each region in the atmosphere to the observed brightness temperature at the given frequency. The peak structure 
of the weighting function gives the range of pressure levels corresponding to the measurements. The weighting function can be 
expressed as a function of microwave opacity of the atmosphere (1). Here, \emph{T$_{b}$} is the brightness temperature, 
\emph{W(p)} is the weighting function as a function of pressure, and \emph{T(p)} is the physical temperature profile of the 
atmosphere.\\

\begin{equation}
T_{b} = \int_{-\infty}^{\infty} W(p)T(p)dlnP
\end{equation}

Fig. 1 shows the relative weighting functions, i.e. weighting function divided by the maximum value of the function, at 0.6 GHz 
and 1.2 GHz with and without alkali metals. In the absence of alkali metals, the relative weighting functions peak at 100 bar and 
30 bar, respectively[\cite{janssen2017mwr}]. At 0.6 GHz, the relative weighting function extends to the deeper atmosphere below 
the 100 bar level, and therefore, the \emph{T$_{b}$} derived using this channel is sensitive to the sources of microwave opacity 
present in the deep atmosphere at \emph{p} greater than 100 bar. The relative weighting function at 0.6 GHz channel shows a broad 
shape with a second maxima at kbar pressure levels which is attributed to the increase in mass absorption coefficients of water 
vapor with pressure. The mass absorption coefficient of ammonia decreases after a maximum near 1000 bar and eventually water 
vapor dominates the opacity in the deep atmosphere. Moreover, the inclusion of free electrons as sources of opacity due to alkali 
metal ionization causes a decrease in the value of the relative weighting function at 0.6 GHz around 100 bar, and a global 
maximum in the relative weighting function emerges at $\sim$ 1 kbar pressure (magenta line). The shift of the global maximum can 
be attributed to the increase in opacity from free electrons with pressure as the ionization fraction of alkali metals increases 
with temperature under thermal equilibrium conditions [\cite{saha1920liii}] (described later in this section). Inclusion of lower 
amounts of alkali metals ([M/H] = -5) will lead to a peak at deeper levels (Fig. 1). However as the metallicity is increased to 
solar, the maximum drifts towards lower pressures around 1 kbar level. This could be attributed to the fact that higher abundance 
of alkali metals can produce higher amount of electrons at relatively lower pressures (magenta line), whereas low abundance of 
alkali metals in Jupiter would need to reach higher pressure ($>$ 1 kbar) to produce equivalent opacity (blue line). Thereby the 
abundance of alkali metals directly affects the shape of weighting function.\\

\begin{figure}[ht!]
\includegraphics[width=0.9\textwidth]{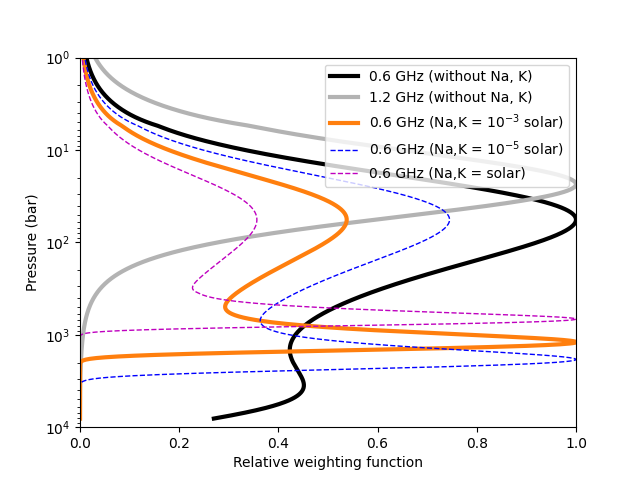}
\caption{Relative weighting functions at 0.6 GHz (black) and 1.2 GHz (gray) for a Jupiter adiabat considering the Hanley model 
[\cite{hanley2009new}] for NH$_{3}$ absorption. The functions peak at 100 bar and 30 bar at 0.6 GHz and 1.2 GHz respectively 
without the inclusion of alkali metals. The inclusion of alkali metals (orange, magenta and blue) decreases the relative 
weighting function at $\sim$ 100 bar and produces a second peak that is observed at $\sim$ 1 kbar pressure due to the opacity 
contributed by free electrons from alkali metal ionization. As the metallicity of alkali metals increase, the global maximum of 
weighting function shifts towards lower pressure.}
\end{figure}

The main sources of microwave opacity at 0.6 GHz and 1.2 GHz are ammonia, water vapor, free electrons, and collision-induced 
absorption by hydrogen and helium. Hydrogen-hydrogen and hydrogen-helium collisions are the dominant sources of collision-induced 
absorption processes in Jupiter. Their magnitude is well constrained due to the invariance of hydrogen and helium abundances in 
Jupiter’s deep atmosphere. The microwave absorption behavior of water and ammonia vapor has been investigated by laboratory 
experiments that show the pressure and temperature dependence of mass absorption coefficients [\cite{devaraj2014centimeter}, 
\cite{karpowicz2011search}, \cite{bellotti2016laboratory}]. In addition, hydrogen, methane, and water vapor contribute to line 
broadening in the ammonia vapor absorption. The models based on laboratory experiments show significant divergent behavior when 
extrapolated to pressures greater than 50 bar and 550 K [\cite{bellotti2016laboratory}]. In order to obtain a robust estimate of  
the range of absorption coefficients at higher temperatures, we test a grid model describing a power scaling relationship with 
temperature based on the \cite{hanley2009new} model of ammonia absorption. For water vapor absorption at microwave frequencies, 
the laboratory models show divergence by orders of magnitude. However, recent laboratory measurements [\cite{steffes2023}] at 
high pressure show that water vapor absorption can be explained by the \cite{bellotti2016laboratory} model. Therefore, 
\cite{bellotti2016laboratory} model is chosen to compute the water vapor opacity which incorporates water opacity measurements at 
high temperatures above 500 K.\\

Free electrons in the atmosphere can act as a source of opacity at microwave wavelengths through the process of free-free 
absorption in which electrons absorb photons during collisions with other ions and electrons. Electrons can be generated by the 
ionization of various elemental and molecular species in the atmosphere. Due to their low ionization energies, alkali metals i.e. 
Na, K are expected to be the major sources of free electrons in the atmosphere [\cite{heays2017photodissociation}]. In Jupiter’s 
atmosphere, the pressure and temperatures corresponding to the transition between the alkali metals and their compounds are 
calculated using an equilibrium cloud condensation model (ECCM) [\cite{atreya1999comparison}, 
\cite{weidenschilling1973atmospheric}] for Jupiter’s adiabat with saturation vapor pressures of Na$_{2}$S and KCl 
[\cite{visscher2006atmospheric}, \cite{morley2012neglected}]. The condensation of alkali metals at solar abundance [Figure 2] 
takes place at 352 bar for KCl and 796 bar for Na$_{2}$S, with corresponding temperatures of 967 K and 1234 K, respectively, 
assuming thermodynamic equilibrium. The condensation of Na$_{2}$S at deeper levels, and a higher solar abundance of Na compared 
to K [\cite{asplund2009chemical}] will cause Na$_{2}$S clouds to be significantly more massive than KCl clouds. Thermochemical 
equilibrium models indicate formation of metal hydrides and hydroxides in gas phase, however they are much lower in abundance 
[\cite{fegley1994chemical}] as compared to the condensates, thereby they will not act as the primary sink of alkali metals in 
Jupiter. Condensation of the alkali metal compounds occurs when the partial pressure of a compound exceeds its saturation vapor 
pressure. If condensation occurs, it causes depletion in the alkali metal abundances at altitudes above the condensation level.\\

\begin{figure}[ht!]
\includegraphics[width=0.9\textwidth]{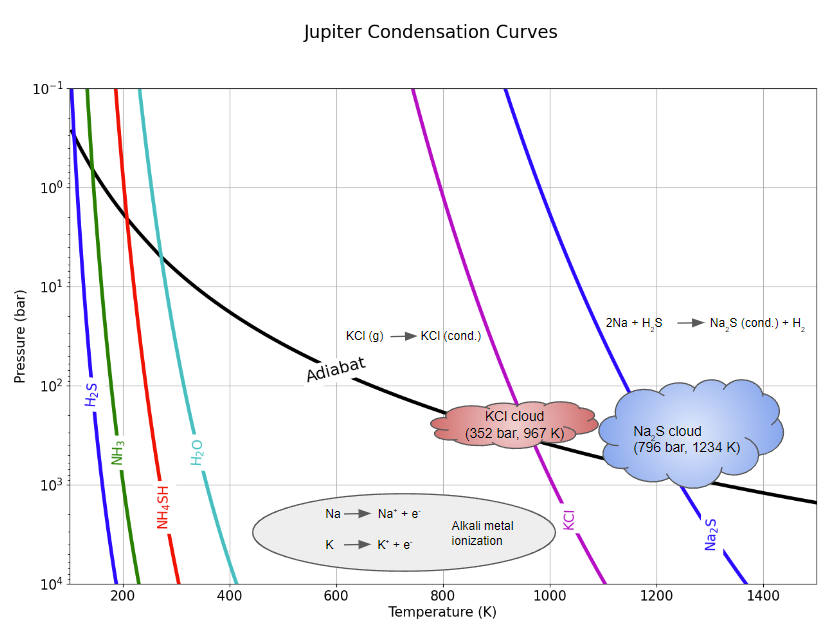}
\caption{Condensation curves of NH$_{3}$, H$_{2}$O, H$_{2}$S and alkali metals Na$_{2}$S and KCl at 1X solar abundance. Our 
calculations are based on the equilibrium cloud condensation model [\cite{atreya1999comparison}], and saturation vapor pressure 
corresponding to Na$_{2}$S and KCl [\cite{visscher2006atmospheric}, \cite{morley2012neglected}]. The cloud bases are at the 
levels where the condensation curves cross the adiabat considering T$_{1bar}$ = 166.1 K [\cite{seiff1998thermal}].
\label{fig:general}}
\end{figure}

At high pressures 100 bar and beyond, alkali metals would undergo ionization to form cold plasma, and the electrons generated in 
the process would act as an additional source of opacity at microwave frequencies. The number density of free electrons due to 
the ionization of alkali metal atoms in the gas phase and is calculated using the Saha ionization [\cite{saha1920liii}] (Eq. 2) 
equation assuming Jupiter’s atmosphere to be in a state of thermal equilibrium. The ionization equation itself assumes a single 
component gas phase system. Thereby, we add the electron densities from ionization of sodium and potassium to determine total 
number density of free electrons. Here, \emph{N$_{e}$} is the electron density, \emph{N} is number density, \emph{$\epsilon$} is 
ionization energy, \emph{$\lambda$} is De Broglie wavelength, \emph{g$_{0}$} and \emph{g$_{1}$} are statistical weights, 
\emph{k$_{B}$} is Boltzmann constant, \emph{m$_{e}$} is mass of the electron and \emph{h} is Planck’s constant.\\

\begin{equation}
\frac{N_{e}^{2}}{N - N_{e}} = \frac{2}{\lambda^{3}}\frac{g_{1}}{g_{0}} e^{-\epsilon/k_{B}T}
\end{equation}

\begin{equation}
\lambda = \sqrt{\frac{h^{2}}{2\pi m_{e} k_{B} T}}
\end{equation}

The brightness temperatures correspond to electromagnetic radiation traveling from the interior of Jupiter radially outwards 
through the atmospheric layers. Thus, the transmission through the deep atmosphere is similar to the transmission through a cold 
plasma medium. The refractive index of microwaves propagating through a cold plasma media can be described by the Appleton-
Hartree equation [\cite{helliwell2014whistlers}]. The formulation is applicable to low-temperature plasma medium both in the 
presence or absence of magnetic fields. At 100-1000 bar pressure levels, the contribution of the magnetic field is insignificant 
in the Appleton-Hartree formulation [\cite{helliwell2014whistlers}]. Therefore, a simplified version of the Appleton-Hartree 
equation (Eq. 4) is used to calculate the complex refractive index of the deep atmosphere using the electron number density 
calculated from the Saha ionization equation. For an unmagnetized cold plasma medium i.e. Jupiter’s deep atmosphere, the Appleton-
Hartree equation is simplified to:\\

\begin{equation}
n^{2} = 1 - \frac{X}{1 - iZ}
\end{equation}

\begin{equation}
\alpha = \frac{2\pi}{\lambda_{ch} Q}
\end{equation}

Here, \emph{X} = $\frac{\omega_{0}^{2}}{\omega^{2}}$, \emph{Z} = $\frac{\nu}{\omega}$  , $\omega_{0}$ is electron plasma 
frequency, $\omega$ is the angular frequency of microwave radiation, $\omega_{h}$ is electron gyro frequency, $\nu$ is electron-
neutral collision frequency, $\lambda_{ch}$ is the frequency of a given MWR channel, \emph{n} is the refractive index, $\alpha$ 
is the extinction coefficient and \emph{Q} is the quality factor i.e. the ratio of squares of real and imaginary parts of the 
refractive index. \\

\subsection{Radiative Transfer Modeling}

In order to draw a comparison between the MWR observations and theoretical knowledge of Jupiter’s atmosphere, a benchmark model 
for the ideal Jupiter atmosphere is constructed using a moist hydrostatic adiabat following the ideal gas law [\cite{li2018high}, 
\cite{li2018moist}]. The specific heat of hydrogen is estimated from the mixing ratio of ortho and para hydrogen assuming thermal 
equilibrium between the ortho and para states. Moreover, the temperature profile of Jupiter’s atmosphere is constructed for two 
cases of reference temperatures: (i) \emph{T} = 166.1 K at the 1-bar pressure level from the Galileo probe 
[\cite{seiff1998thermal}] and (ii) \emph{T} = 168.8 K at the 1-bar pressure level based on the reanalysis of the Voyager radio 
occultation experiment at Jupiter [\cite{gupta2022jupiter}].  Ammonia and water vapor are considered vapors for the moist adiabat 
and their partial pressure is controlled by the cloud condensation process by forcing the partial pressures to be equal to their 
saturation vapor pressures. In the deep atmosphere of Jupiter, water and ammonia are not expected to form clouds; however, alkali 
metals are expected to undergo condensation. Therefore, a similar approach is applied to alkali metals to estimate the 
concentration of alkali metals present in the gas phase available for the ionization process.\\

Spectral radiance is proportional to the physical temperature of the atmosphere in the Rayleigh-Jeans limit. For microwave 
frequencies, we compute the brightness temperature (\emph{T$_{b}$}) from the physical temperature using Eq. (1). The opacity of 
Jupiter’s atmosphere is the sum of opacities from individual sources discussed in the previous section i.e. ammonia, water, free 
electrons, and collision-induced absorption. The abundances of ammonia and water vapor have been assumed to be 2.7 and 5 times 
the solar abundance [\cite{li2020water}, \cite{li2017distribution}]. Because there is no a priori information on the alkali metal 
abundance in Jupiter, Therefore, we compare two cases, one without alkali metals (baseline) and another with alkali metals 
(treatment) in order to provide a comparison between our current knowledge of Jupiter and MWR data.\\

The spatial resolution of MWR data also provides the limb darkening coefficient at six microwave frequencies. Limb darkening 
(\emph{L$_{d}$}) is defined as the percent change in \emph{T$_{b}$} at a given viewing angle relative to \emph{T$_{b}$} at a 
position looking vertically down to the planet center i.e. nadir. For our simulations, we compute the limb darkening at a 45-
degree angle from the nadir. The MWR channels at 0.6 GHz and 1.2 GHz are chosen to provide a comparison between theory and 
observations at higher pressures using \emph{T$_{b}$} and \emph{L$_{d}$} as the observables for comparison. The benchmark case of 
the ideal Jupiter atmosphere is compared with MWR observations as a function of latitude between -40 and 40 degrees 
planetocentric latitude. Data from higher latitudes are neglected due to the presence of signatures from synchrotron radiation 
that is inseparable from the atmospheric contribution. \\

\begin{figure}[ht!]
\includegraphics[width=0.9\textwidth]{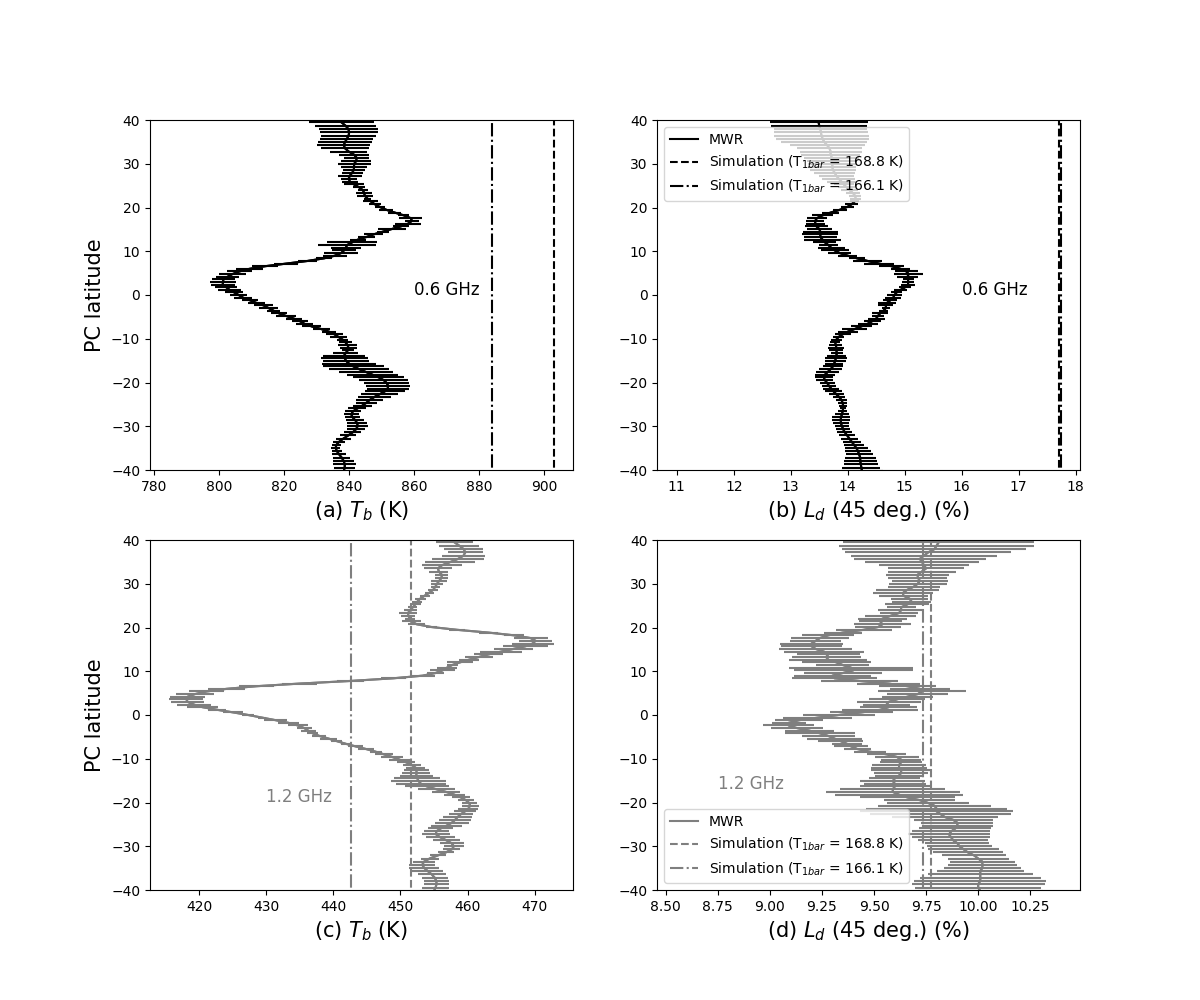}
\caption{Limb darkening and brightness temperature MWR observations compared with simulation results at 0.6 GHz and 1.2 GHz 
corresponding to Jovian adiabats at (i) \emph{T$_{1bar}$} = 166.1 K  and  (ii) \emph{T$_{1bar}$} =  168.8 K , (a) \emph{T$_{b}$} 
vs. latitude at 0.6 GHz, (b) \emph{L$_{d}$} vs. latitude at 0.6 GHz, (c) \emph{T$_{b}$} vs. latitude at 1.2 GHz, (d) 
\emph{L$_{d}$} vs. latitude at 1.2 GHz.
\label{fig:general}}
\end{figure}

A latitudinal variation in brightness temperatures is observed at both 0.6 and 1.2 GHz (Figure 3, panels (a) and (c)). The small-
scale variations in \emph{T$_{b}$} and \emph{L$_{d}$} in all the panels can be attributed to variations in the atmospheric 
temperature structure and composition. It is important to note that the baseline case (without alkali metals) corresponds to two 
different temperature profiles of Jupiter's atmosphere for two different \emph{T$_{1bar}$}. There is an agreement between the 
baseline case and observations at 1.2 GHz in the equatorial region (panel (c)). On the other hand, brightness temperatures at 0.6 
GHz are lower than the baseline case by 40-60 K at all latitudes (panel (a)) indicating the possibility of an additional source 
of opacity. Such a source is also supported by a depressed  \emph{L$_{d}$} observed by MWR; it is 4 percent less than the 
\emph{L$_{d}$} magnitude of the ideal Jupiter atmosphere across all latitudes (panel (b)). The mismatch between the  baseline and 
observations at 0.6 GHz is much greater than the uncertainty in measurements and variations in \emph{T$_{b}$} and \emph{L$_{d}$}. 
Since the brightness temperatures correspond to different pressure regions in the atmosphere, the anomalous observations at 0.6 
GHz must be attributed to the presence of an additional opacity source in the deep atmosphere or to a different opacity source 
that absorbs more effectively at 0.6 GHz than at 1.2 GHz. We test four confounding factors: (1) the distribution of ammonia, (2) 
the ammonia opacity at temperatures exceeding the range of laboratory measurements, (3) the opacity of water at high temperatures 
and (4) the contribution of alkali metals. The theoretical brightness temperature and limb darkening at 0.6 GHz and 1.2 GHz is 
shown in Fig. 3. \\

The latitudinal distribution of brightness temperatures and limb darkening from the forward model indicates the decrease in limb 
darkening from the equator to the pole at 0.6 GHz. It is opposite to the variation of limb darkening at 1.2 GHz across the 
latitudes. This effect could be attributed to the free electrons in the deep atmosphere which could be inferred from the shift in 
the contribution functions toward higher pressures in presence of alkali metals (Fig. 1). Alkali metals greatly affect the 
absorption behavior at 0.6 GHz which dominates the effect of gravitation on limb darkening.\\

\subsection{Ammonia, Water and Alkali Metals}

Brightness temperature variations with latitude and the spectral inversion of brightness temperatures show a non-uniform 
distribution of ammonia vapor in Jupiter’s atmosphere in the deep atmosphere region [\cite{li2017distribution}, 
\cite{ingersoll2017implications}]. Therefore, the non-uniform distribution of ammonia could contribute to variations in microwave 
opacity of the deep atmosphere. In order to estimate the effect of ammonia concentration variations, we perturb the ammonia 
profile in the model and use a scaling factor to vary the magnitude of ammonia vapor concentration in the model as described in 
Eq. (6). \\

\begin{equation}
q_{NH_{3}}(P) =q_{NH_{3}, 0}(P) -(q_{NH_{3},0}(P) - q_{NH_{3}, MWR}(P))s  
\end{equation}

Here, \emph{q$_{NH_{3}}$} is the ammonia mass mixing ratio at a given pressure \emph{P}, \emph{q$_{NH_{3},0}(P)$} is the 
homogeneous ammonia mixing ratio which is set to 2.7 times solar abundance for NH$_{3}$ $\sim$ 360 ppm 
[\cite{li2017distribution}] from the deep atmosphere till the NH$_{3}$ vapor saturation point. Above the saturation point, the 
mixing ratio follows the NH$_{3}$ saturation vapor pressure curve. \emph{q$_{NH_{3}, MWR}$(P)}is the mixing ratio retrieved from 
MWR inversion. We use a scaling factor to vary the ammonia mixing ratio between the homogeneous case to MWR derived profiles. The 
scaling factor, s ranges from 0 to 1.5 where 0 is the case for homogeneous mixing ratio. Increasing s to 1 will change the 
ammonia profile to MWR inversion case for equator and mid-latitude regions. We also extend the scaling factor to 1.5, in order to 
take into account the low ammonia mixing ratio observed at the North Equatorial Belt (NEB) of Jupiter 
[\cite{li2017distribution}].\\

NH$_{3}$ opacity measurements are currently not available for high temperatures (~ 550 K-3000 K) corresponding to Jupiter’s deep 
atmosphere and there is a decrease in the magnitude of absorption of NH$_{3}$ at high pressures. Thereby, we invoke a scaling 
factor to the NH$_{3}$ absorption coefficient to provide an estimation of the opacity at high temperatures. The mass absorption 
coefficient of ammonia is estimated by multiplying the temperature-scaling law to the absorption coefficient based on 
\cite{hanley2009new} (Eq. 7). In this equation, $\alpha$ is the absorption coefficient of NH$_{3}$, h is the opacity factor, 
\emph{T} is temperature and \emph{T$_{c}$} is reference temperature equal to 750 K. The NH$_{3}$ opacity models show that the 
absorption coefficient peaks at 750 K and decreases at temperatures beyond 750 K. In the simulations, the scaling factor is 
multiplied to the NH$_{3}$ opacity at temperatures higher than \emph{T$_{c}$}. The power law index (h) is varied from 1 to 5 
keeping the ammonia concentration constant, i.e., 2.7 times solar abundance. We also keep the water vapor constant at 5 times solar abundance as the laboratory measurements demonstrate that water vapor absorption does not show a significant increase with pressure and can be said to be relatively transparent when compared to the previous model of microwave absorption [\cite{steffes2023}].\\

\begin{equation}
\alpha(NH_{3}) \sim \left(\frac{T_{c}}{T}\right)^{h}
\end{equation}

Changing the ammonia profile and introducing the additional temperature-dependent scaling factor produce brightness temperature and limb darkening divergent from MWR data at 0.6 GHz as shown in Figure 4a. The difference between \emph{T$_{b}$} from the model and observations is in the range of 50-200 K at 0.6 GHz. Reducing the ammonia concentration causes a monotonic increase in \emph{T$_{b}$} and a decrease in \emph{L$_{d}$}. Further, reducing the ammonia opacity shows a similar trend in \emph{T$_{b}$}, while a saturation in \emph{L$_{d}$} is expected at a power law factor of 5. Changing the ammonia profile and ammonia opacity has a similar effect on \emph{T$_{b}$} and \emph{L$_{d}$} at 1.2 GHz. However, overall, the variation in the MWR observations at 1.2 GHz can be explained by these two factors and does not require the inclusion of alkali metals. The 1.2 GHz observations correspond to $\sim$ 20 bar (Fig. 1), much above the cloud base of alkali metals and at relatively lower pressure levels. Therefore, the contribution of free electrons to opacity is expected to be less due to lower temperatures, and the opacity contribution of ammonia vapor dominates at 1.2 GHz. However, a comparison of MWR observations at both frequencies clearly implies that the variation in ammonia vapor opacity cannot solely explain the anomalous observations at the 0.6 GHz channel.//

\begin{figure}
\centering
\includegraphics[width=0.9\textwidth]{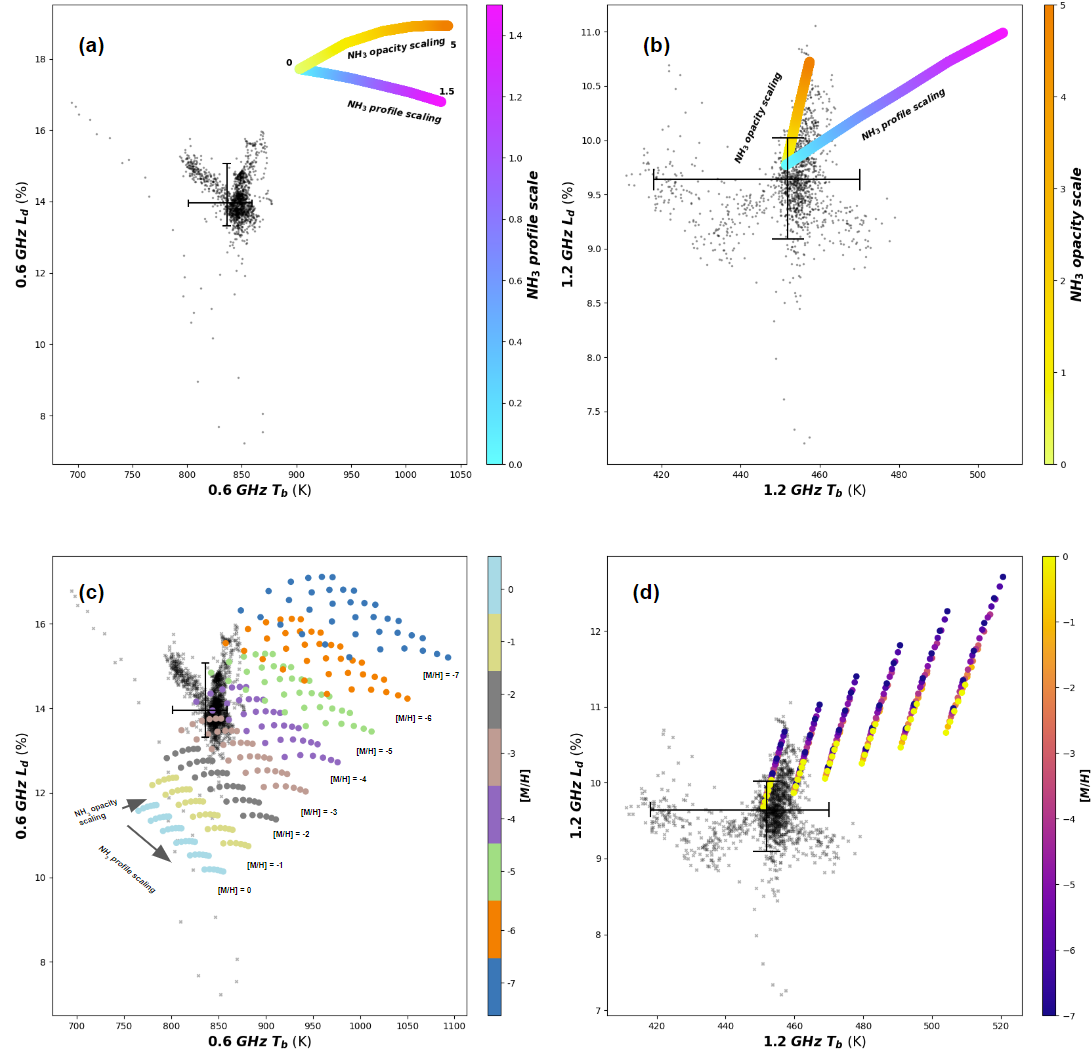}
\caption{Comparison is drawn between the Juno MWR observations and the results of the radiative transfer model for \emph{T$_{b}$} and \emph{L$_{d}$} at 0.6 GHz and 1.2 GHz, keeping the water abundance constant $\sim$ 5 times solar abundance. (a, b) Jupiter’s atmosphere in the absence of alkali metals with only variations in the NH$_{3}$ vapor profile and the NH$_{3}$ opacity, (c, d) Jupiter’s atmosphere in the presence of alkali metals with variations in the NH$_{3}$ vapor profile and the NH$_{3}$ opacity. The NH$_{3}$ profile of Jupiter's atmosphere is varied using a scale from 0 to 1.5 to take into account the contribution of non-uniform distribution of NH$_{3}$ vapor observed by MWR [\cite{li2017distribution}]. NH$_{3}$ opacity at temperatures above 750 K undergoes power law scaling as a function of atmospheric temperature (Eq. 7). In the absence of alkali metals, the changes in NH$_{3}$ vapor profile and the scaling in NH$_{3}$ vapor opacity deviate significantly from Juno MWR observations at 0.6 GHz. However, in the presence of alkali metals of low metallicity, i.e., in the range of -2 to -5, there is an agreement between model results and MWR observations. Observations at 1.2 GHz can be explained by variations in the NH$_{3}$ vapor profile and the NH$_{3}$ opacity independent of opacity contributions from alkali metals.}
\label{fig:general}
\end{figure}

Fig. 4c, 4d examines the overall effect of alkali metals and ammonia vapor on the \emph{T$_{b}$} and \emph{L$_{d}$} at 0.6 GHz and 1.2 GHz. We vary the alkali metal metallicities in a range from 0 to -7 (solar abundance of Na and K according to \cite{asplund2009chemical}) for each condition of NH$_{3}$ profile scaling and NH$_{3}$ opacity scaling. The volume mixing ratio of Na and K corresponding to abundance in solar photosphere [\cite{asplund2009chemical}] are 3.46 x 10$^{-6}$ (\emph{[Na/H]} = -5.76) and 2.14 x 10$^{-7}$ (\emph{[K/H]} = -6.97), respectively. Therefore, we simulate a wide range of ammonia opacity conditions for a given alkali metal abundance (colored dots). Both NH$_{3}$ profile and opacity scaling cause a change in \emph{T$_{b}$} and \emph{L$_{d}$}, which is shown by the annotation in the figure. The variation in \emph{T$_{b}$} and \emph{L$_{d}$} is similar to the pattern in Fig. 4a. NH$_{3}$ profile scaling causes a decrease in \emph{L$_{d}$}, while the scaling in NH$_{3}$ vapor opacity causes \emph{L$_{d}$} to increase at 0.6 GHz. For each case of metallicity, we then perform a scaling in ammonia vapor and ammonia opacity as described previously in this section. This provides us with a matrix of \emph{T$_{b}$} and \emph{L$_{d}$} to take into account all possible sources of opacity, i.e., collision-induced absorption, ammonia, water vapor, and free electrons from alkali metals. The free electron opacity is calculated from the Hartree-Appleton equation explained in the previous section.\\

\begin{figure}[ht!]
\includegraphics[width=0.9\textwidth]{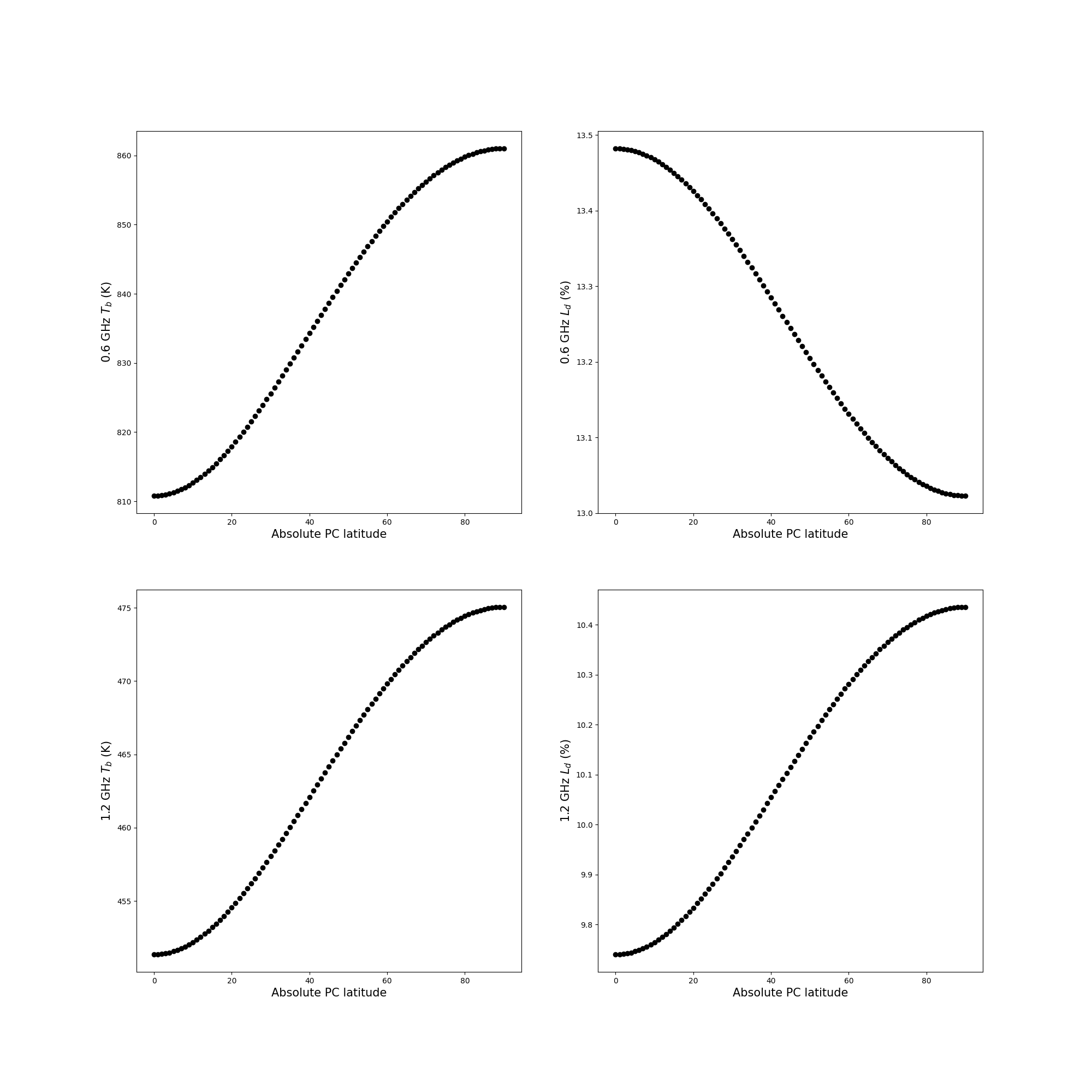}
\caption{Latitudinal variation of brightness temperature and limb darkening of Jupiter’s atmosphere at 0.6 GHz and 1.2 GHz at [M/H] = -3}
\label{fig:general}
\end{figure}

When we compare the new model result with MWR observations (Fig. 4b), we observe that model matches with observations at 0.6 GHz 
for free electrons corresponding to alkali metal metallicities in the range of -2 to -5 (chocolate colored patches), i.e. 
10$^{-2}$ to 10$^{-5}$ times the solar abundance. There is an agreement between the model and observations at 1.2 GHz for the 
same range of metallicities. The addition of free electrons from alkali metals dominates the effect of gravity (Fig. 5) and we 
expect the limb darkening to decrease from equator to the poles assuming uniform mixing ratio of water and ammonia vapor. It 
serves as a baseline to understand the sole effect of free electrons on latitudinal variation of microwave radiation from 
Jupiter’s deep atmosphere. \\

\section{Discussions} \label{sec:style}

We infer metallicity of the alkali metals in Jupiter to be much lower than the solar value. A possible indication of low 
metallicity of the alkali metals in a hot Jupiter exoplanet was first proposed by \cite{demory2011high} as one plausible 
explanation for the high albedo of Kepler-7b. They derived an alkali metal abundance 10-100 times lower than the solar value. 
Since then the abundance of alkali metals has been derived for several other giant exoplanets, with abundances ranging from  
$\sim$ 100 times below solar to $\sim$ 100 times above solar, although the uncertainties are large. Recent observations of two 
hot Jupiters or Saturns with clear or mostly clear atmospheres were made. The alkali metal abundance for one such hot Jupiter 
(HAT-P-1b) is found to be sub-solar [\cite{chen2022detection}], while it was found to be solar to greatly super-solar for the 
other (WASP-96b)[\cite{nikolov2022solar}]. Considering the relatively small sample size of hot Jupiters with clear atmospheres, 
it is premature to make a meaningful comparison between their alkali metal metallicity and the metallicity in Jupiter presented 
in this paper. On the other hand, it is instructive to compare the abundance of alkali metals in Jupiter from this work with the 
abundance of the other heavy elements. While the opacity contribution from alkali metals suggest that Na and K are strongly 
depleted relative to solar at the level probed by MWR at 0.6 GHz, all other heavy elements are enriched by a factor of 
approximately three to five; while nitrogen is highly variable but enriched, and the water abundance remains uncertain 
[\cite{atreya2019origin}, \cite{li2020water}, \cite{li2017distribution}, \cite{mahaffy2000noble}]. The comparison to other heavy 
metal measurements from the Galileo probe corresponds to much lower pressures i.e. $<$ 22 bars.  The estimation of alkali metal 
metallicity from MWR implies lower metallicity at much higher pressures. The results (Fig. 4b) provide an important constraint on 
alkali metal abundance at pressures sensitive to 0.6 GHz channel. A [M/H] = -1 for alkali metals provides too much opacity while 
too little abundance or absence of alkali metals does not provide sufficient opacity to match the MWR observations at 0.6 GHz.\\

The low abundance of alkali metals indicated by MWR observations could be attributed to any of the following scenarios. (i) 
Initially enriched alkali metals, consistent with the other heavy elements in the atmosphere, are depleted by chemical reactions 
with other constituents deep in the atmosphere, resulting in a low abundance of Na and K at  $\sim$ 1 kilobar level sufficient to 
provide the free electrons to explain the MWR data at 0.6 GHz. Fegley and Lodders [\cite{fegley1994chemical}] predict, for 
example, the formation of gas-phase species of Na and K in the atmosphere i.e. NaCl, NaOH, and KOH. Should there be chemical 
mechanisms that could selectively deplete K in the atmosphere, leaving Na to be the most significant contributor to free 
electrons in the deep atmosphere, the metallicity of Na would be expected to be in the range of 0 to -2 i.e. solar to highly sub-
solar abundance (Appendix B).  (ii) Unconventional planet formation processes, whereby Jupiter did not accrete a solar complement 
of alkali metals, or that the alkali metals are not well mixed at greater depths. If the depletion of alkali metals at $\sim$ 1 
kbar inferred in this paper is representative of their bulk abundance, it could be indicative of the depletion of all rock-
forming elements, with significant implications for the formation and evolution of Jupiter. Our conclusion of depletion is based 
on the data of the 0.6 GHz channel, whose weighting function peaks at 1 kilobar level with the inclusion of alkali metals. Thus, 
we are confident about the result only at this level. Alkali metals could well be more abundant deeper in the atmosphere and they 
could have been depleted by some as yet unknown mechanism before reaching the 1 kilobar level though the degree of depletion 
would have to be huge. \cite{barshay1978chemical} considered one such possibility, where silicates were found to be a way of 
sequestration of gas phase alkali metals. However, a later study by \cite{fegley1994chemical} found it to be an ineffective 
mechanism. Further modeling and laboratory studies are needed to cover the full parameter space of combined thermochemistry of 
alkali metal and rock cloud forming species corresponding to the very high temperature and high pressure conditions of the deep 
atmosphere of Jupiter, together with any dynamical effects, before drawing any firm conclusions about depletion of alkali metals 
in bulk Jupiter below the level to which the MWR data of this paper are sensitive.\\
 
The new constraints on the abundance of alkalis are linked to their low ionisation potential, and the fact that the electrons 
that they provide directly affect opacities at 0.6 and 1.2 GHz (see Eq. 4). But when present, they are strong absorbers at 
visible wavelengths (e.g., \cite{burrows2000near}) and therefore directly affect the planetary radiative flux. The low abundances 
that we derive imply that a radiative zone may be present in Jupiter [\cite{guillot1994nonadiabatic}, 
\cite{guillot2004interior}]. Interestingly, this could explain at the same time the relatively low abundance of CO observed in 
Jupiter’s atmosphere compared to expectations for a fully convective deep atmosphere [\cite{cavalie2023subsolar}].\\

\section{Software and third party data repository citations} \label{sec:cite}
The software for the radiative transfer package will be available at zenodo archive (\url{https://doi.org/10.5281/zenodo.7893914
}) and the MWR data used in this work, and associated files for data visualization are available at archive (
\url{https://doi.org/10.5281/zenodo.7893817}). They can be made available upon request\\

\software{High-performance Atmospheric Radiation Package (HARP) [\cite{li2018high}, \cite{bhattacharya_ananyo_2023_7893914}]}

\counterwithin{figure}{section}
\renewcommand{\thefigure}{\Alph{section}.\arabic{figure}}
\appendix
\section{Appendix A: Electron Density and Conductivity}

The electron density of Jupiter’s atmosphere is governed by two fundamental processes: (i) condensation of alkali metal 
condensates i.e. Na$_{2}$S and KCl, and (ii) ionization of alkali metals in thermal equilibrium. Fig. 2 shows the pressure levels 
corresponding to the cloud base of Na$_{2}$S and KCl based on their saturation vapor pressures. Cloud condensation reduces the 
amount of alkali metals available in gas phase that act as a source of free electrons, and restricts the abundance of Na and K 
corresponding to their respective saturation vapor pressure. In the cloud region, electron density is controlled by saturation 
vapor pressure of alkali metals whereas below the cloud base, electron densities are governed by metallicity of alkali metals. 
Thereby, it is evident that condensation controls the electron density and thereby, conductivity at low pressure levels. 
Condensation limited ionization is observed at low pressure (below 1 kbar) irrespective of the alkali metal abundance as the 
electron density lines converge (Fig. A.1 (a)). Fig. A.1 (a) and (b) show the presence of a kink in electron density and their 
respective conductivity at the cloud base corresponding to different alkali metal abundances. However, condensation does not play 
a significant role in governing the electron densities at $\sim$ 1 kbar pressure level corresponding to the global maxima in the 
weighting function at 0.6 GHz (Figure 1). \\

\setcounter{figure}{0} 
\begin{figure}[ht!]
\centering
\includegraphics[width=0.7\textwidth]{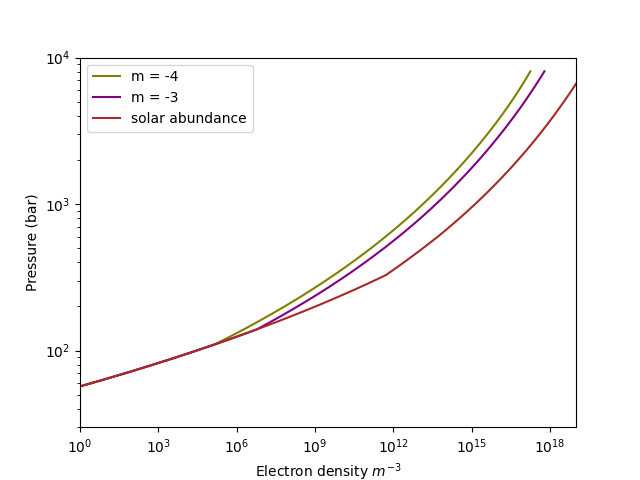}\\
\textbf{(a)}\\

\includegraphics[width=0.7\textwidth]{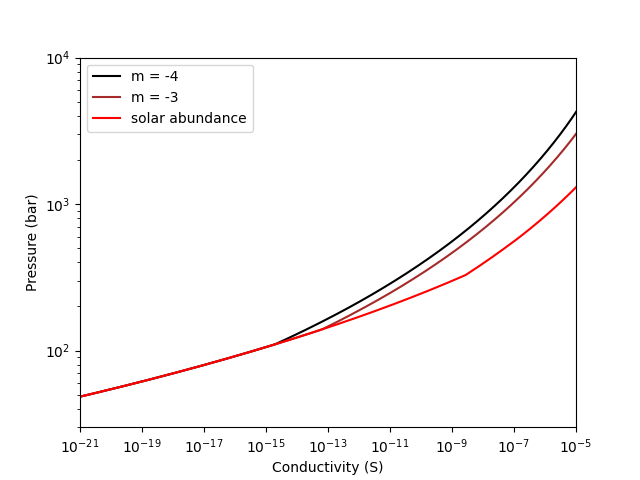}\\
\centering
\textbf{(b)}\\

\caption{(a) Electron density of Jupiter’s deep atmosphere at the solar abundance and \emph{[M/H]} = -3 and -4, (b) electrical conductivity of Jupiter’s deep atmosphere at the solar abundance and [M/H] = -3 and -4. 
\label{fig:appendix_figure}}
\end{figure}

The electron density of the deep atmosphere is much lower than in the case of alkali metals at solar abundance. It is the true representation of the electron density of the deep atmosphere. At greater pressures, hydrogen behaves as a semiconductor and becomes the major contributor to the electron density [\cite{LIU2008653}]. The electrical conductivity of the atmosphere is calculated using Drude’s equation. It provides an estimate of the conductivity due to the free electrons provided by alkali metal ionization.\\

\section{Appendix B: Selective Depletion of Alkali Metals}

Even though Na$_{2}$S has a deeper condensation level compared to KCl, the cloud condensation is governed by atmospheric temperature, and does not reflect the chemical reactivity of alkali metals. K is more electropositive than Na and thereby, is expected to be more reactive as compared to Na. Therefore, it is possible that there can be a chemical mechanism that could selectively deplete K into other compounds, leaving Na as the only source of free electrons in Jupiter. Under such conditions, we find that Na metallicity should be in the range of 0 to -3 to match the MWR observations. The increase in alkali metal metallicities can be attributed to two factors: (i) low ionization energy of K, and (ii) Na$_{2}$S condenses much below KCl (Figure 2). Thereby, a larger amount of Na is required to produce enough free electrons to match the MWR brightness temperatures and limb darkening.\\
\setcounter{figure}{0} 
\begin{figure}[htbp]
\includegraphics[width=0.9\textwidth]{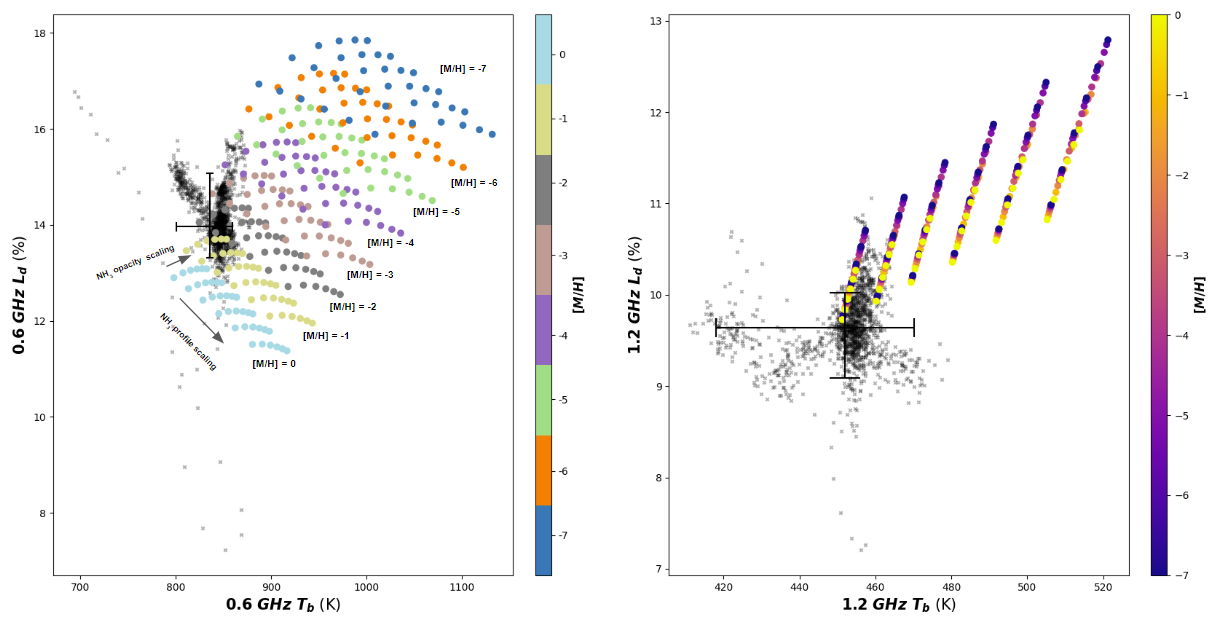}
\caption{Limb darkening and brightness temperature comparison of MWR observations and forward model results at 600 MHz and 1.2 GHz for metallicities ranging from 0 to -7 at different ammonia vapor concentration profiles and opacities. It showcases the sole effect of free electrons due to the ionization of Na, without considering any contribution from K.}
\label{fig:appendix_figure}
\end{figure}

The elimination of K from the atmosphere highlights the role of the elemental abundance of Na required to match the MWR observations. The results of the forward model in Fig. B.1 indicate the possible solutions of Na metallicity under different conditions of ammonia vapor concentration profiles and microwave opacities. It is observed that the range of Na metallicity is expected to be from 0 to -3 i.e. solar abundance to highly sub-solar abundance. Thus, metallicity of Na required is expected to be higher than those considering both Na and K to be sources of free electrons.\\

\section{Appendix C: Jovian Adiabats and Comparison of MWR with High Temperature Adiabat}

Fig. 2 shows that brightness temperatures at 600 MHz from two adiabats differ by approximately 15 K. The relative weighting function for the adiabats is that of the ideal Jupiter’s atmosphere without the inclusion of opacity due to free electrons from alkali metals. It shows a peak at $\sim$ 100 bar. From the difference in physical temperature of the atmosphere of the two adiabats, it is seen that the difference reaches $\sim$ 10-15 K at 100 bar level (Fig. C.1). The weighting function at 600 MHz also extends below 100 bar which could explain the difference in brightness temperatures. An interesting observation is that the difference in adiabat temperatures increases with increase in atmospheric pressure. This increase can be attributed to the temperature dependent specific heat of the atmospheric constituents.\\
\setcounter{figure}{0} 
\begin{figure}[ht!]
\includegraphics[width=0.9\textwidth]{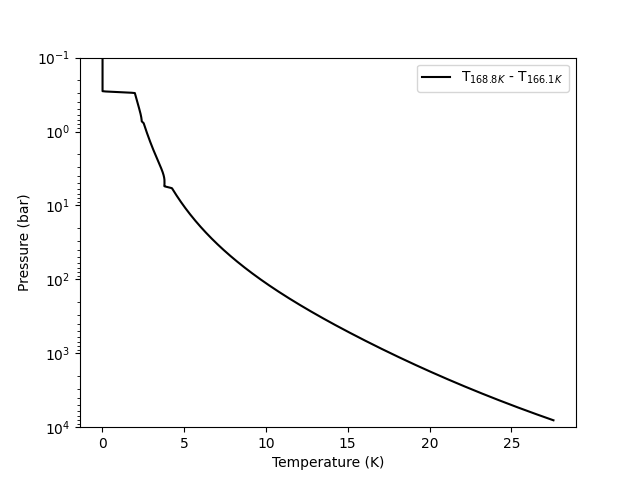}
\caption{Pressure v. temperature difference in temperatures of Jovian adiabats constructed using T$_{1bar}$ = 168.8 K and T$_{1bar}$ = 166.1 K
\label{fig:appendix_figure}}
\end{figure}

The interior models of Jupiter generally use a high temperature in the range of 170-180 K at the outer boundary (1 bar pressure level) [\cite{gupta2022jupiter}, \cite{miguel2022jupiter}]. These temperatures are about 10-15 K higher than the measurements from the Galileo probe (166.1 K)[\cite{seiff1998thermal}] and Voyager radio occultation reanalysis (168.8 K)[\cite{gupta2022jupiter}]. A simulation of brightness temperatures and limb darkening at 0.6 GHz and 1.2 GHz is carried out for all cases of alkali metal metallicities, ammonia concentration and opacity variation assuming \emph{T$_{1bar}$} = 175 K. It can be clearly seen in Fig. C.2 that high temperature at 1 bar doesn't match  with entire range of MWR observations for both the frequencies. Some alternate possibilities could be the presence of a non-adiabatic gradient or a radiative layer in Jupiter’s deep atmosphere that can possibly account for a higher temperature at 1 bar level. However, the mismatch with MWR at 1.2 GHz poses a serious question on the assumption. The current measurements of temperature at 1 bar level are from limited radio occultation experiments. There is a need for radio science experiments from equator to the poles, in order to estimate the true variability in temperatures at 1 bar.\\

\begin{figure}[ht!]
\includegraphics[width=0.9\textwidth]{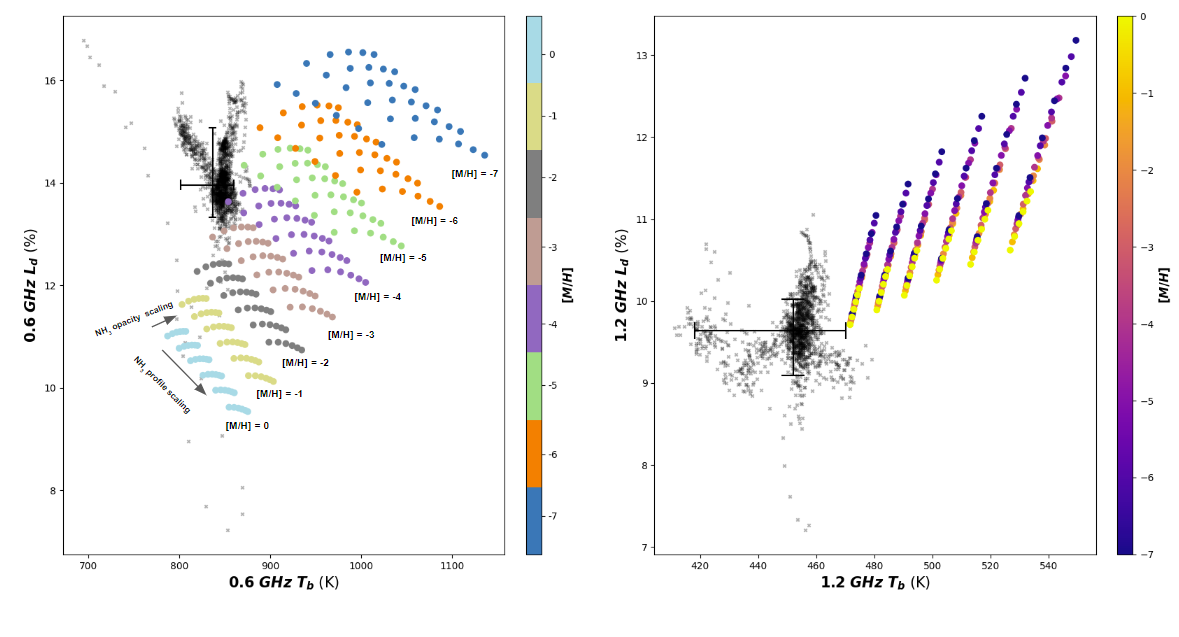}
\caption{Limb darkening and brightness temperature comparison of MWR observations and forward model results at 600 MHz and 1.2 GHz for metallicities ranging from 0 to -7 at different ammonia vapor concentration profiles and opacities considering T$_{1bar}$ = 175 K
\label{fig:appendix_figure}}
\end{figure}


\bibliography{sample631}{}
\bibliographystyle{aasjournal}



\end{document}